\documentclass[aps,prl,floatfix,twocolumn,superscriptaddress,showpacs,10pt]{revtex4-1}
\usepackage{amssymb, amsmath,color,mciteplus,graphicx,subfigure}
\usepackage{calc,epsfig,epstopdf,color,mciteplus,bm,mathrsfs}
\usepackage{times}
\usepackage{float}
\usepackage{lipsum}

\begin{document}

\title{Robust and Fast Quantum State Transfer on Superconducting Circuits}

\author{Xiao-Qing Liu}
\affiliation{Guangdong Provincial Key Laboratory of Quantum Engineering and Quantum Materials,
and School of Physics and Telecommunication Engineering,
South China Normal University, Guangzhou 510006, China}

\author{Jia Liu}\email{liuj.phys@foxmail.com}
\affiliation{Guangdong Provincial Key Laboratory of Quantum Engineering and Quantum Materials,
and School of Physics and Telecommunication Engineering,
South China Normal University, Guangzhou 510006, China}
\affiliation{Guangdong-Hong Kong Joint Laboratory of Quantum Matter, and Frontier Research Institute for Physics, South China Normal University, Guangzhou 510006, China}

\author{Zheng-Yuan Xue}\email{zyxue83@163.com}
\affiliation{Guangdong Provincial Key Laboratory of Quantum Engineering and Quantum Materials,
and School of Physics and Telecommunication Engineering,
South China Normal University, Guangzhou 510006, China}
\affiliation{Guangdong-Hong Kong Joint Laboratory of Quantum Matter, and Frontier Research Institute for Physics, South China Normal University, Guangzhou 510006, China}
\begin{abstract}
Quantum computation attaches importance to high-precision quantum manipulation, where the quantum state transfer with high fidelity is necessary. Here, we propose a new scheme to implement the quantum state transfer of high fidelity and long distance, by adding on-site potential into the qubit chain and enlarging the proportion of the coupling strength between the two ends and the chain. In the numerical simulation, without decoherence, the transfer fidelities of 9 and 11 qubit chain are 0.999 and 0.997, respectively. Moreover, we give a detailed physical realization scheme of the quantum state transfer in superconducting circuits, and discuss the tolerance of our proposal against decoherence. Therefore, our scheme will shed light on quantum computation with long chain and high-fidelity quantum state transfer.

\end{abstract}
\date{\today}

\maketitle

Quantum computation can solve problems effectively that are beyond the capacity of classical computers \cite{QC,shor}. High-precision quantum manipulation, a fundamental requirement for quantum computation, has been implemented in different physical systems such as trapped ions \cite{trapion1,trapion2,trapion3,trapion4}, quantum dots \cite{qdot}, photons in optics \cite{photon1,photon2,photon3}, superconducting quantum circuits \cite{squbit1,squbit2}, etc. The transfer of quantum states from one location to another is a way in quantum manipulation, and high-fidelity quantum state transfer (QST) is essential to both quantum communication and large-scale quantum computation \cite{transfer1,transfer2}. There are two main ways to achieve QST: long-distance QST and short-distance QST. Long-distance QST is an essential ingredient of quantum networks \cite{ld}, and it is achieved by interfacing between stationary quantum states and flying qubits \cite{ld1,ld2,ld3,ld4}. Short-distance QST is of great importance for on-chip quantum information processing \cite{Bose}, and it is achieved by exchange interactions among qubits. Short-distance QST can be achieved mainly by two methods. The first approach is to achieve the transport of states through a series of exchanges \cite{swap1,swap2}. However, in solid-state systems, this exchange is usually achieved by elaborately adjusting the time-dependent magnetic field, which is not friendly to experimental implementation, and the process introduces a large amount of noise, which makes fidelity worse. The spin chain is another method to achieve state transmission, which can effectively avoid the above shortcomings.

\begin{figure}[tbp]
  \centering
  \includegraphics[width=1\linewidth]{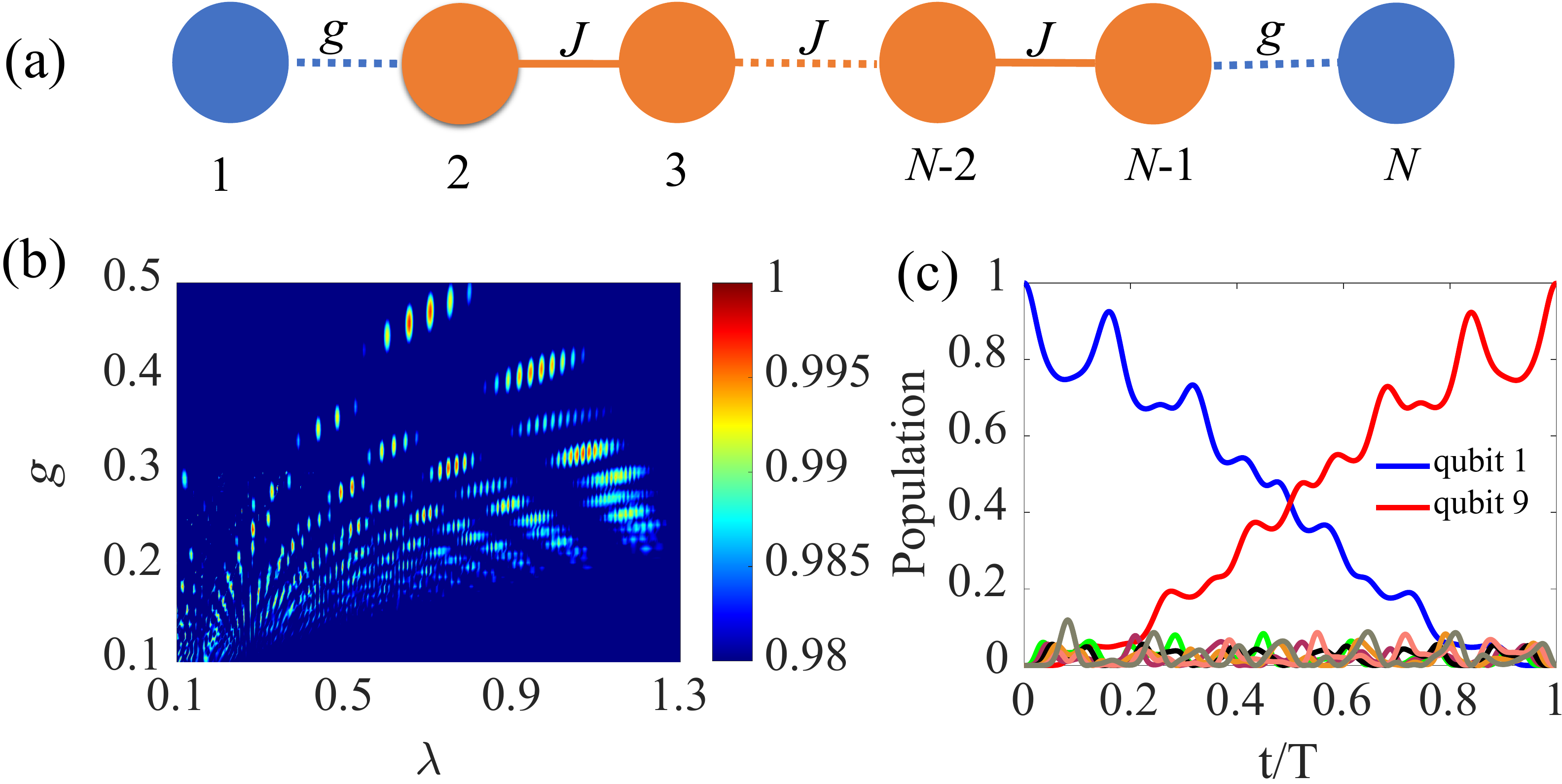}
  \caption{(a) An illustration of the scheme, a spin chain links the sender (site 1) and the receiver (site \emph{N}) with coupling strength \text{\emph{g}}, and the coupling strength of the interchain is $\emph{J}$, which is set to be 1 in the numerical calculation. (b) The numerical simulation of large-scale search results in the case \emph{N}=9. The horizontal coordinate is the absolute value of on-site potential, and the vertical is the coupling strength of both ends. The diagram represents the fidelity distribution on the \text{\emph{g}}-$\lambda$ plane. (c) The specific process of state transfer as an example of \textcolor{blue}{9 spin qubits}. The system prepares the initial state $|1\rangle_1$. After a period of evolution time $\tau$, the state is retrieved from the extreme of the chain with confidence.} \label{Fig1}
\end{figure}

\begin{figure*}[tbp]
  \centering
  \includegraphics[width=0.9\linewidth]{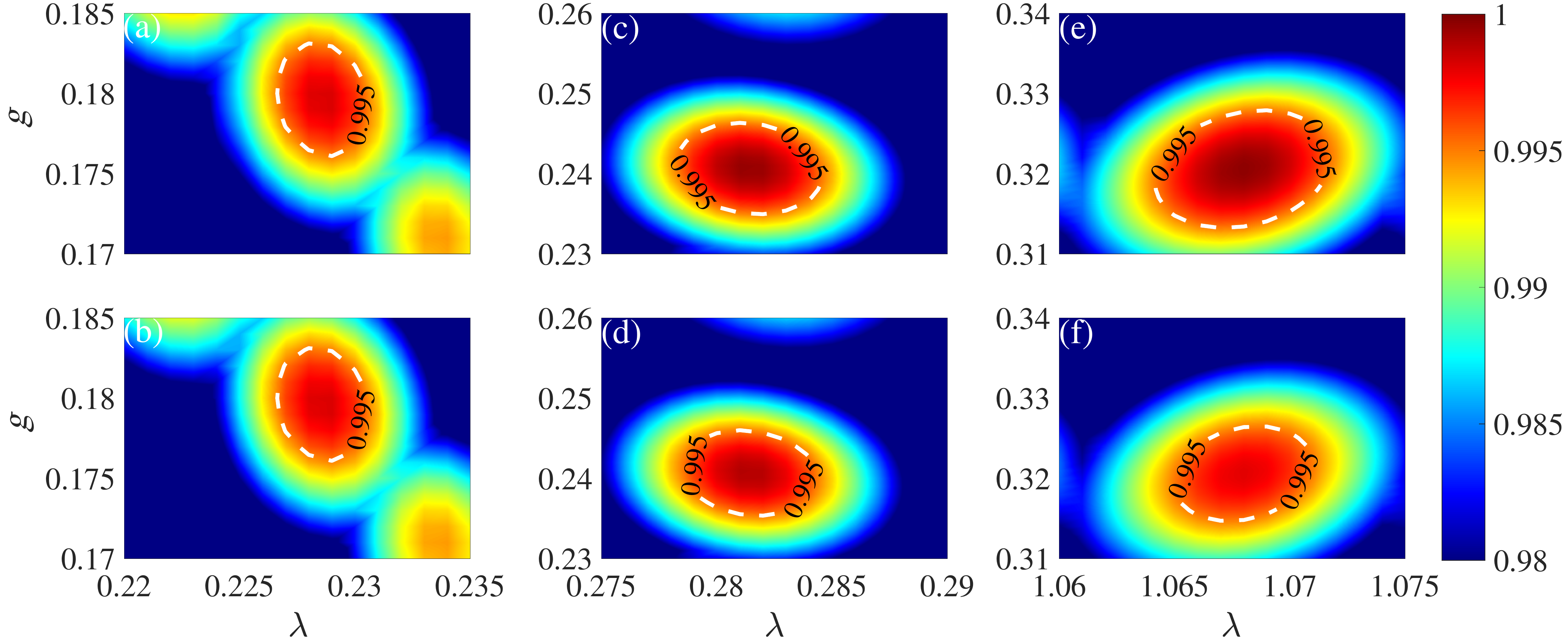}
  \caption{The fidelity distribution in the plane of ~$\text{\emph{g}}$ and ~$\lambda$ for \textcolor{blue}{9 spin qubits}, most parts in the region have fidelity higher than 0.99. The fidelity without decoherence is represented in Figs. (a), (c) and (e). The case with decoherence is shown in Figs. (b), (d) and (f) where the ratio of decoherence rate $\Gamma$ to the coupling strength $\emph{J}$ is $10^{-3}$.}\label{Fig2}
\end{figure*}

The Quantum spin chain for QST was first proposed by Bose \cite{Bose}. Since the idea that the spin chain can be used as a short-range transmission came up, the advantages of the spin chain, natural evolution on its own and no interfaces, have attracted the interests. However, the fidelity of QST becomes worse as the chain is longer, which is obviously not in line with the actual requirements. In order to achieve perfect transmission with long distance, many kinds of programs have been presented \cite{net, inversion1, inversion2, network, graph, dual, receiver, endgate, boundary}. And it is shown that the precise control of coupling strength is a good solution \cite{net, network, inversion1}, but it is challengeable for experiment operations. Therefore, to make the experimental operation as simple as possible, the idea of reducing the ratio of the coupling intensity between the two ends and the chain is put forward \cite{odd, endcoupling}, where the whole system is simplified into a two-state or three-state system, and then using the characteristics of the energy spectrum, the state is resonately transfered \cite{odd, Yao}. It is worth noting that in this scheme, high efficiency is accompanied by a small ratio, if high fidelity is obtained, then the transmission time closely associated with the coupling strength will be very long, at the same time, the influence of external factors on the system will enlarge. Therefore, the question of how to shorten the time of state transmission and obtain higher fidelity is our objective.

In this paper, we find another method to achieve state transmission with shorter time and higher fidelity. The idea is to enlarge the proportion of the coupling strength between the two ends and the chain and add alternating on-site potential, which can improve transmission efficiency. The numerical simulation results indicate that our plan can not only meet the requirements of high fidelity, but also have certain robustness for external interference. Meanwhile, the priority of the superconducting quantum circuits is used to demonstrate the plan.

The protocol of description is shown in Fig. \ref{Fig1}(a), where the coupling strength inside the spin chain is \emph{J}, and the two relative ends of the chain are connected by the coupling strength $\text{\emph{g}}$. Thence, the Hamiltonian of the system can be described by
 \begin{eqnarray}
 \label{EqH}
 \mathcal{H}_1=&&\sum^{N}_{j=1}\lambda_{j} S_j^{z}+\sum^{N-2}_{j=2}J( S_j^{+}S_{j+1}^{-}+\mathrm{H.c.})\notag\\
 &&+\text{\emph{g}}(S_1^{+}S_2^{-}+S_{N-1}^{+}S_{N}^{-}+\mathrm{H.c.}),
 \end{eqnarray}
where $S^{\pm}=(S^x\pm\textrm{i}S^y)$, $S^z=S^{+}S^{-}-I/2$, $\lambda_j=(-1)^j\lambda$ is the on-site potential, and ~$\emph{N}$ is the total number of spin qubits.
With Jordan-Wigner transformation \cite{Jorder-Wigner}, i.e., $c_{j}=\exp[\textrm{i}\pi(\sum^{j-1}_{l=0}S_l^{+}S_l^{-})]S_{j}^{-}$, the Hamiltonian is transformed to
\begin{eqnarray}
\label{EqH1}
\mathcal{H}_1^{T}=&&\sum^{N}_{j=1}\lambda_{j} c_j^{\dagger}c_j+\sum^{N-2}_{j=2}J(c_j^{\dagger}c_{j+1}+\mathrm{H.c.})\notag\\
&&+\text{\emph{g}}(c_1^{\dagger}c_2+c_{N-1}^{\dagger}c_{N}+\mathrm{H.c.}).
\end{eqnarray}
With the help of `operators' commutation relations, we find $[\mathcal{H}_1^{T},\sum_{j=1}^{N}n_j]=0$, where $n_j=c_j^{\dagger}c_j$ is the `particle' number operator, and thus the Hamiltonian of the Eq. (\ref{EqH}) is transformed to another Hilbert space consisting of eigenvectors of particle number operators. For a spin qubit, spin-down and spin-up states are adopted and represented by $|0\rangle$ and $|1\rangle$, respectively, and then the single excitation subspace of the system is spanned by $|1\rangle_j=|0,0...010...0\rangle$, which denotes the $j$th qubit spins up while others spin down.

\begin{figure*}[tbp]
\centering
\includegraphics[width=0.9\linewidth]{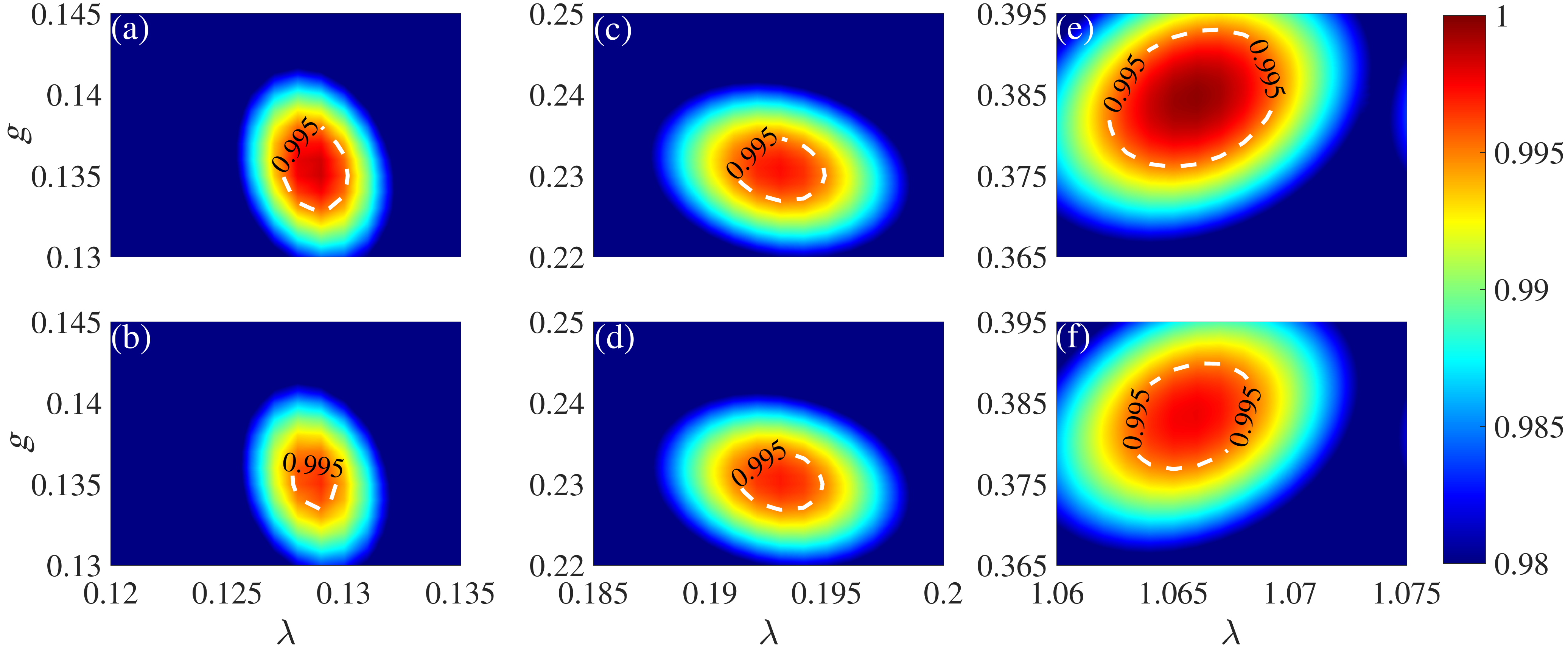}
\caption{The fidelity distribution in the plane of ~$\text{\emph{g}}$ and ~$\lambda$ for \textcolor{blue}{11 spin  qubits}, most parts in the region have fidelity higher than 0.99. The fidelity without decoherence is represented in Figs. (a), (c) and (e). The case with decoherence is displayed in Figs. (b), (d) and (f) where the ratio of decoherence rate $\Gamma$ to the coupling strength $J$ is $10^{-3}$.}\label{Fig3}
\end{figure*}
Next, the process of QST can be specifically described as follows. The system prepares the initial state $|\psi(0)\rangle=|1\rangle_1$. After a period of evolution time $\tau$, the system will evolve into a superposition of a series of states, i.e., $|\psi(\tau)\rangle=\sum_{j=1}^N C_j |1\rangle_j$, where $C_{j}$ is normalization factor. As a measurement of the merit of state transmission, the fidelity of QST can be represented as $F=\lvert C_N \rvert^2$, and also calculated by $ F= \text{Tr}(\rho_{\tau}\rho_{N})$ \cite{QC}, where $\rho_{\tau}=|\psi(\tau)\rangle\langle\psi(\tau)|$ is the density operator at the end and $\rho_{N}=|1\rangle_N\langle1|$ is the the ideal result. The solution of ${\rho_{\tau}}$ is based on the Lindblad master equation through a differential method, i.e.
\begin{eqnarray}
\label{EqHL1}
\partial{\rho}/\partial{t}=-\textrm{i}[\mathcal{H}^T_1,\rho]+\frac{\Gamma}{2}[\mathcal{L}(a_1)+\mathcal{L}(a_2)],
\end{eqnarray}
where $\mathcal{L}(\mathcal{A})=2\mathcal{A}\rho\mathcal{A}^{\dagger}-\mathcal{A}^{\dagger}\mathcal{A}\rho-\rho\mathcal{A}^{\dagger}\mathcal{A}$, $a_1=\sum_{j=1}^{N}c_{j}$ and $a_2=\sum_{j=1}^{N}n_{j}$ are the corresponding decay and dephasing of the whole system respectively, and $\Gamma$ is the decoherence rate. The investigation will show the influence of decoherence on our scheme.

In our proposal, there are three parameters $\{\emph{J},\text{\emph{g}},\lambda\}$, which together determine the performance of QST. Our goal is to realize the QST with a longer chain and higher fidelity by searching for appropriate parameters. As we all know, the transfer quality is related to length \cite{Bose} and parity of the chain \cite{odd}. In our numerical simulation, under the condition of the same parameter range, odd-numbered qubits have better performance for QST than even-numbered qubits. Therefore, we first display the QST of \textcolor{blue}{9 spin qubits} with our scheme. Here and after, setting $ \emph{J}=1 $, we plot the effect of parameters $\{\text{\emph{g}},\lambda\}$ for QST, as shown in Fig. \ref{Fig1}(b). We found that the fidelity of QST in some parameter regions can be used to realize quantum computation, and then some representative parameter regions are selected and enlarged, as shown in Fig. \ref{Fig2}, where the first and second row of the figure correspond to the numerical results without and with decoherence, respectively.

In previous works \cite{odd,Yao}, high-fidelity transmission can be achievable under the restriction condition $\emph{J} \gg \emph{\text{g}}$, but evolution time in such relation to the coupling coefficient is very long. If you consider the system's decoherence, the fidelity of the whole operation will be discounted. In our proposal, adding on-site potential to the qubits in the chain, even though $\text{\emph{g}}$ takes a slightly larger value, we can get high-efficiency QST. We find that for the case $N=9$, the value of $\text{\emph{g}}$ is between 0.15 and 0.35, transfer of high fidelity is still complete. The results are shown in Fig. \ref{Fig2}, where the fidelity of QST without decoherence in Fig. \ref{Fig2}(a), (c) and (e) can reach 0.999 and the fidelity with decoherence in Fig. \ref{Fig2}(b), (d) and (f) is greater than 0.997. Our proposal not only realizes the state transmission of high fidelity, but also relaxes the restriction and reduces the evolution time. More importantly, the scheme is relatively easy to implement in the experiment, which will be discussed in detail later. From what has been discussed above, the introduction of on-site potential is of great advantage.

Subsequently, we extend the length of the chain to 11 and search for suitable parameters for achieving high-fidelity QST. We scan in the $\text{\emph{g}}-\lambda$ plane, and find that in the region $\text{\emph{g}}=0.1\sim0.5$ and $\lambda=0.1\sim1.5$, results with high fidelity can be gotten. The region with high fidelity is selected for magnification, the enlarged picture is shown in Fig. \ref{Fig3}. The partial fidelity is greater than 0.997 when there is no decoherence involved as shown in Fig. \ref{Fig3}(a), (c) and (e). The partial fidelity is more than 0.996 when interaction with the outside is considered as shown in Fig. \ref{Fig3}(b), (d) and (f). In order to check the length our proposal can achieve, we also explore the case of a chain with \textcolor{blue}{13 spin qubits} and find that the best result is 0.998 without decoherence and 0.997 with decoherence. If the length continues to increase, the effect will be worse even if there is no decoherence involved, that is the longer the chain, the lower the fidelity of the transfer will be.

Compared with previous work, even if the coupling intensity \text{\emph{g}} on the edge is slightly larger, we can obtain better fidelity. We provide a qualitative explanation from the perspective of layout. For $N=9$, the layout of $|1\rangle_j$ where $j$ is from 2 to 8, is very small as shown in Fig. \ref{Fig1}(c), even if they are summed up, the result is still small, so the high-fidelity QST is completed through a large detuning. The same processing method is also applicable to the length of 11, the difference is that the sum of layout $|1\rangle_j$ that is from 2 to 10 is sometimes 0.5, so high-fidelity QST is completed through resonance.

With the theoretical results above, we continue to show that a spin chain can be simulated on superconducting circuits, where adjacent transmons are coupled by a large capacitance. The Hamiltonian of the coupled system can be written as
\begin{eqnarray}
\label{EqH2}
\mathcal{H}_2=\sum_{l=1}^{N}\frac{\omega_{l}}{2}\sigma_{l}^{z}+\sum_{l=1}^{N-1}\Omega_{l}(\sigma_{l}^{\dagger}\sigma_{l+1}+\mathrm{H.c.}),
\end{eqnarray}
where $\omega_{l}$ and $\sigma_{l}^{\dagger}$ denote the frequency and the creation operator for the $l$th transmon, respectively, and $\Omega_{l}$ is coupling strength between ${l}$th and $(l+1)$th transmons. In accordance with the scheme, tunable coupling strength is necessary and can be achieved by our parametric modulation on the qubits \cite{modulation1,modulation2,modulation3,modulation4}. Specifically speaking, we adopt frequency driving for the $l$th transmon except the first in the form of $\omega_{l}(t)=\omega_{l}+\varepsilon_{l}(t)$, with $\varepsilon_{l}(t)=\dot{F}_{l}(t)$, where $F_{l}(t)=f_{l}\sin[\omega_{l}^{d}t +\phi_{l}^{d}(t)]$ and $\omega_{l}^{d}$ is the frequency of the driving \cite{modulation5}. To see the effectiveness of modulation more clearly, the transformation is conducted by virtue of the unitary operator
\begin{eqnarray}
U(t)\!=\!\exp \left \{\!-\frac{\textrm{i}}{2}\bigg (\!\sum_{l=1}^{N}(\omega_{l}\!-\!m_{l})t\sigma_{l}^{z}+\sum_{l=2}^{N}\!F_{l}(t)\sigma_{l}^{z}\!\bigg)\right \},\notag\\
\end{eqnarray}
 where $m_{l}$ is a parameter and determined according to the Hamiltonian to be simulated \cite{detuning1,detuning2}. And transform the Hamiltonian in Eq. (\ref{EqH2}) into an interaction picture as
\begin{eqnarray}
\label{EqH2I}
\mathcal{H}_2^T(t)=&&\textrm{i}\dot{U}^{\dagger}(t)U(t)+U^{\dagger}(t)\mathcal{H}_2U(t) \notag\\
=&&\sum_{l=1}^{N}\frac{m_{l}}{2}\sigma_{l}^{z}+\Omega_{1}(e^{-\textrm{i}\Delta_{1}t}e^{-\textrm{i}F_{2}(t)}\sigma^{\dagger}_{1}\sigma_{2}+\mathrm{H.c.}) \notag\\
&&+\sum_{l=2}^{N-1}\Omega_{l}(e^{-\textrm{i}\Delta_{l}t}e^{\textrm{i}F_{l}(t)}e^{-\textrm{i}F_{l+1}(t)}\sigma_{l}^{\dagger}\sigma_{l+1}+\mathrm{H.c.}),\notag\\&&
\end{eqnarray}
where $\Delta_{l}\!=\!\omega_{l+1}\!-\!\omega_{l}\!-\!m_{l+1}\!+\!m_{l}$, $F_{l}(t)$ can be changed by virtue of the Jacobi-Anger identity $e^{\textrm{i}z\cos\theta}=\sum_{-\infty}^{\infty}\textrm{i}^{n}J_{n}(z)e^{\textrm{i}n\theta}$, where $\emph{J}_{n}(z)$ is the \emph{n}th Bessel functions of the first kind. When $\Delta_{l}=\omega_{l+1}^d$ for odd $l$ and $\Delta_{l}=-\omega_{l+1}^d$ for even $l$ are satisfied, effective Hamiltonian can be obtained through using the rotating-wave approximation and neglecting the oscillating terms, i.e.
\begin{eqnarray}
\label{EqH2I}
\mathcal{H}_2^{eff}=\sum_{l=1}^{N}\frac{m_{l}}{2}\sigma_{l}^{z}+\sum_{l=1}^{N-1}\Omega_{l}^{eff}(\sigma_{l}^{\dagger}\sigma_{l+1}+\mathrm{H.c.}),
\end{eqnarray}
where
 \begin{equation}
 \begin{aligned}
   \begin{array}{lr}
    \Omega_{1}^{eff}=\Omega_{1}J_{1}(f_{2})e^{\textrm{i}(\phi_{2}^{d}+\pi)}, & \hbox{$l$=1;} \\
     \Omega_{l}^{eff}=\Omega_{l}J_{0}(f_{l})J_{1}(f_{l+1})e^{\textrm{i}(\phi_{l+1}^{d}+\pi)}, & \hbox{$l$ is odd;} \\
     \Omega_{l}^{eff}=\Omega_{l}J_{0}(f_{l})J_{1}(f_{l+1})e^{-\textrm{i}\phi_{l+1}^{d}}, & \hbox{$l$ is even.}
   \end{array}
 \end{aligned}
 \end{equation}
On the basis of Eq. (\ref{EqH2I}), our plan can be achieved by choosing $m_{l}=-\lambda$ when ${l}$ is odd, and $m_{l}=\lambda$ when ${l}$ is even. It is noticed that $\omega_{l+1}^{d}=\Delta_{l}$ can be expressed in another form $m_{l+1}-m_{l}=\omega_{l+1}-\omega_{l}-\omega_{l+1}^{d}$, the right side of which represents frequency distance between neighboring transmons and external driving, that is detuning, so potential distribution can be obtained by adjusting detuning for $2\lambda$ upward or downward in the experiment. Effective coupling strength $\Omega_{l}^{eff}$ can be achieved through suitable $\Omega_{l}$ and external driving amplitudes $f_{l}$ and $ f_{l+1}$ to make $\Omega_{1}^{eff}=\Omega_{N-1}^{eff}=\text{\emph{g}}$, $\Omega_{l}^{eff}=J$ set up. As mentioned above, we can change the frequency and amplitude of driving to obtain adjustable on-site potential and coupling strengths as we plan above.

\begin{figure}[tbp]
  \centering
  \includegraphics[width=1\linewidth]{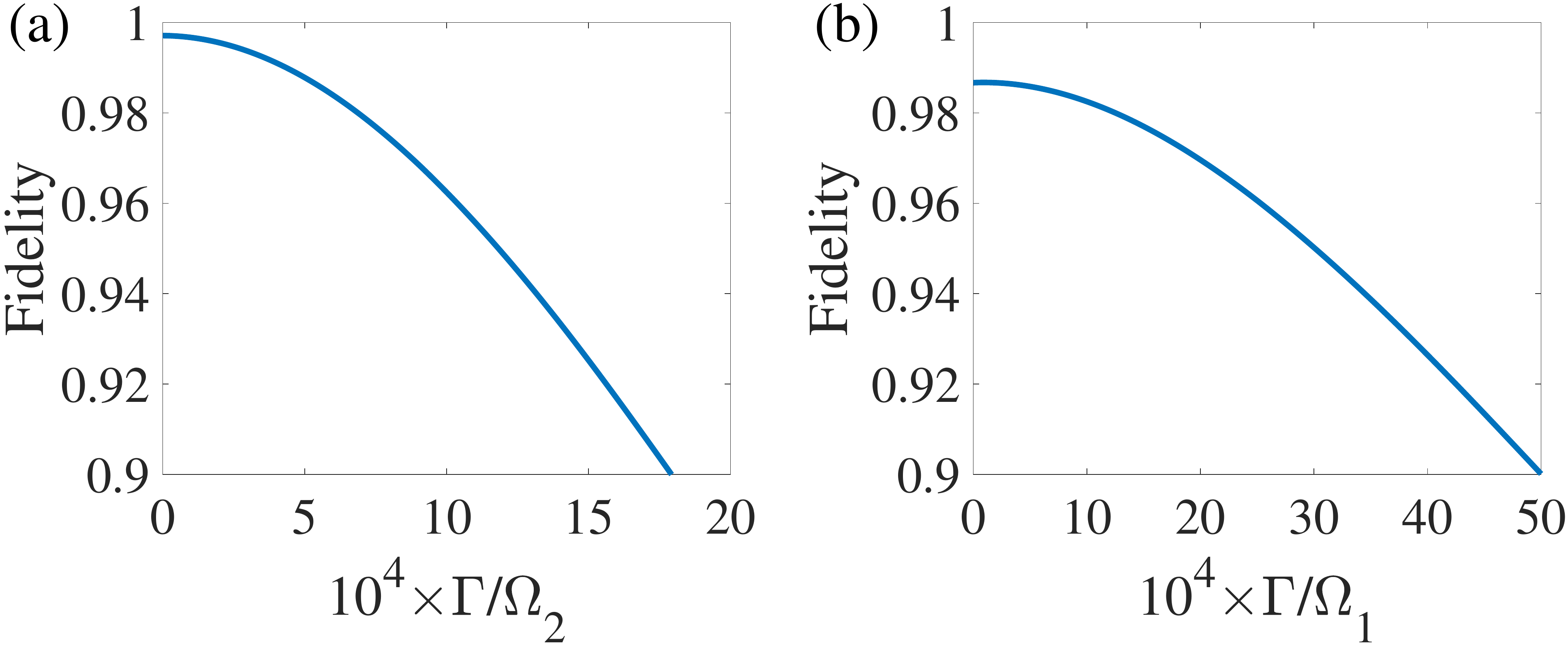}
  \caption {The influence of decoherence $\Gamma$ on the fidelity of quantum state transfer on superconducting circuits. The left figure (a) is the chain length of 9, coupling strength $\Omega_2$ equals to $2\pi\times15$ MHz and the right (b) is 11, $\Omega_1$ corresponds to $2\pi\times10$ MHz.}\label{Fig4}
\end{figure}

We proceed to illustrate QST of different lengths in superconducting quantum circuits in the case of maximum fidelity as examples. In the case of length for $N=9$, the coupling strengths of the two ends and the middle are the same separately, that is $\Omega_1=\Omega_8$, $\Omega_2=\Omega_3=...=\Omega_7$, and there is a quantitative relation of strengths between the middle and edges, the optimal of which is $\Omega_1=0.31\Omega_2$. As for the frequency of external driving, it is increasing with the interval $\Delta$, for example $\omega_3^{d}=\omega_2^{d}+\Delta$ and $\omega_4^{d}=\omega_3^{d}+\Delta$, besides, the form of initial frequency $\omega_2^{d}$ and $\Delta$ is individually $\omega_2^{d}=12.11\Omega_2$ and $\Delta=\Omega_2$, and $\Delta_{l}$ equals to $\omega_{l+1}^{d}$ to induce time-modulation interaction. There is an expression between the amplitude $f_l$ and $\Omega_l$, that is $\Omega_1J_1(f_2)=0.241n$, $\Omega_lJ_0(f_l)J_1(f_{l+1})=n$ when $l$ is taken from 2 to 7, $\Omega_8J_0(f_7)J_1(f_8)=0.241n$ where \emph{n} is random, so the amplitude $f_l$ can be obtained by inverse Bessel function. The absolute value of last parameter $m_l$ can be satisfied the equation $|m_l|=0.08\Omega_2$.

And for $N=11$, we adopt another approach to the coupling strengths among transmons that they are identical and remarked as $\Omega_1$, and choose the driving frequency in the same way as above. And driving frequency, frequency distance and the absolute value of potential are expressed as $\omega_2^{d}=56.8\Omega_1$, $\Delta=5.8\Omega_1$, $|m_l|=0.69\Omega_1$, respectively. The driving amplitude $f_l$ can be obtained by inversely solving equation $\Omega_1J_1(f_2)=0.384n$, $\Omega_1J_0(f_l)J_1(f_{l+1})=n$ from $l=2$ to 9, $\Omega_1J_0(f_{10})J_1(f_{11})=0.384n$ where \emph{n} is random.

It's worth noting that no matter what $\Omega_2$ for $N=9$ or $\Omega_1$ for $N=11$ is, the fidelity of state transfer doesn't change as long as the parametric equations above are satisfied. Here, we take $\Omega_2=2\pi\times15$ MHz for a demonstration with respect to 9 transmons and $\Omega_1=2\pi\times10$ MHz for a demonstration about 11 transmons, and results in the Fig. \ref{Fig4} display the tolerance to decoherence. With the decoherence rate $\Gamma=2\pi\times5$ kHz, numerical simulation on the left shows the transmission from the first transmon to the last is with fidelity of 99.36$\%$, and on the right is 98.62$\%$, which shows our scheme is robust against decoherence.

In summary, we propose a scheme to implement high-fidelity QST in a spin chain. Our proposal is achieved by modulating coupling strength and adding on-site potential, which effectively reduces the evolution time and has strong robustness against decoherence. \textcolor{blue}{In other words, the benefit of our scheme is robust against the decoherence due to shorter time brought by coupling strength}. Meanwhile, our proposal proves to be realized in superconducting quantum circuits. That long distance and high fidelity quantum state transmission will shed light on the realization of a quantum computer in the future.

\begin{acknowledgements}
We thank Pu Shen for the helpful discussion. This work is supported by the Key-Area Research and Development Program of Guangdong province (Grant No.~2018B030326001), the National Natural Science Foundation of China (Grant No.~12275090 and No.~11904111), the Project funded by China Postdoctoral Science Foundation (Grant No.~2022M711222), and Guangdong Provincial Key Laboratory (Grant No. 2020B1212060066).
\end{acknowledgements}

\end{document}